\documentclass[letterpaper, 10 pt, conference]{ieeeconf}
\IEEEoverridecommandlockouts
\usepackage{graphicx}
\usepackage{microtype}
\usepackage{amsmath,amssymb}
\usepackage{siunitx}
\usepackage{booktabs}
\usepackage{cite}
\usepackage{url}
\usepackage{xcolor}
\usepackage{tikz}
\usetikzlibrary{arrows.meta,fit,backgrounds}
\usepackage[hidelinks]{hyperref}

\newif\ifanon
\anonfalse

\makeatletter
\newcommand{\citeurl}[2]{\expandafter\gdef\csname citeurl@#1\endcsname{#2}}
\def\lcite@nolink{\@ifnextchar[{\lcite@nolink@}{\lcite@nolink@[]}}
\def\lcite@nolink@[#1]#2#3#4{#4}
\newcommand{\lcite@one}[1]{\@ifundefined{citeurl@#1}{\cite{#1}}{\hyper@linkurl{\cite{#1}}{\csname citeurl@#1\endcsname}}}
\newcommand{\lcite}[1]{\begingroup
  \let\hyper@@link\lcite@nolink
  \def\lcite@sep{}\@for\lcite@k:=#1\do{\lcite@sep\lcite@one{\lcite@k}\def\lcite@sep{,}}\endgroup}
\makeatother
\citeurl{durre2006igra}{https://journals.ametsoc.org/view/journals/clim/19/1/jcli3594.1.xml}
\citeurl{ecmwf2016radiosonde}{https://doi.org/10.21957/cblxtg}
\citeurl{spectrum2026buoy}{https://spectrum.ieee.org/ai-weather-forecasting-balloons-buoys}
\citeurl{spisak2026sensors}{https://arxiv.org/abs/2602.02714}
\citeurl{windborne2023gdas}{https://ams.confex.com/ams/103ANNUAL/meetingapp.cgi/Paper/415803}
\citeurl{bellemare2020loon}{https://doi.org/10.1038/s41586-020-2939-8}
\citeurl{gossamer2024prop}{https://doi.org/10.2514/1.C037905}
\citeurl{drela2009ddwfttw}{https://www.boatdesign.net/attachments/ddw2-pdf.28167/}
\citeurl{sachs2005albatross}{https://doi.org/10.1111/j.1474-919x.2004.00295.x}
\citeurl{loyd1980crosswind}{https://doi.org/10.2514/3.48021}
\citeurl{erhard2013kites}{https://doi.org/10.1109/TCST.2012.2221093}
\citeurl{nock2007winged}{https://doi.org/10.2514/1.31922}
\citeurl{yoder2019sails}{https://doi.org/10.2514/1.C035353}
\citeurl{yoder2021tethered}{https://doi.org/10.2514/1.C036057}
\citeurl{sushko2025skysail}{https://patents.google.com/patent/US12466586B2/en}
\citeurl{cherubini2015awe}{https://doi.org/10.1016/j.rser.2015.07.053}
\citeurl{lambert2003aerostat}{https://doi.org/10.2514/2.3149}
\citeurl{bird2014soaring}{https://doi.org/10.2514/6.2014-0263}
\citeurl{valbal_paper}{https://doi.org/10.1109/AERO.2017.7943912}
\citeurl{chandrasekhar1943}{https://doi.org/10.1103/RevModPhys.15.1}
\citeurl{mahony2008complementary}{https://doi.org/10.1109/TAC.2008.923738}
\citeurl{astrom2021feedback}{https://press.princeton.edu/books/hardcover/9780691193984/feedback-systems}
\graphicspath{{figures/}}

\title{\LARGE \bf
Skysail: Altitude Control of Long-Duration Weather Balloons by Extracting Work from Wind Shear
}

\ifanon
\author{Anonymous authors\thanks{\rule{0pt}{2.6\baselineskip}}}
\hypersetup{pdfauthor={}}
\else
\author{Michal Adamkiewicz$^{1}$, Andrey Sushko$^{1}$%
\thanks{$^{1}$The authors are with WindBorne Systems, Palo Alto, CA, USA,
        {\tt\small \{michal, andrey\}@windbornesystems.com}. Corresponding author: Michal Adamkiewicz.}%
}
\hypersetup{pdfauthor={Michal Adamkiewicz, Andrey Sushko}}
\fi

\begin{document}
\maketitle
\thispagestyle{empty}
\pagestyle{empty}

\begin{abstract}
Long-duration weather balloons collect repeated atmospheric profiles by venting lifting gas and dropping ballast, but these finite consumables limit the number of profiles each balloon can obtain. We present an altitude-control system that extracts mechanical work from atmospheric wind shear. A small wing suspended 450 m below a zero-pressure balloon experiences relative airflow from the wind velocity difference between the two altitudes. Controlling the wing's pitch produces aerodynamic forces that drive ascent and descent. We develop a model for sizing the system, a lightweight aerodynamic governor for tether deployment, and a control architecture for altitude holding and repeated profiling. Flight experiments conducted over two years accurate control of vertical motion. During a 43-hour interval, the system completed eleven profiles and accumulated 53 km of vertical travel without releasing ballast or venting gas. A separate eight-day flight completed 24 profiles exceeding 1 km in vertical extent, accumulating 159 km of vertical travel. Control authority depends on the available wind shear, with conventional ballast and vent actuation retained for trim and fallback control. These results demonstrate that atmospheric wind shear can sustain repeated balloon profiling while reducing dependence on consumable actuators.
\end{abstract}

\begin{figure}[t]
  \centering
  \includegraphics[width=\columnwidth]{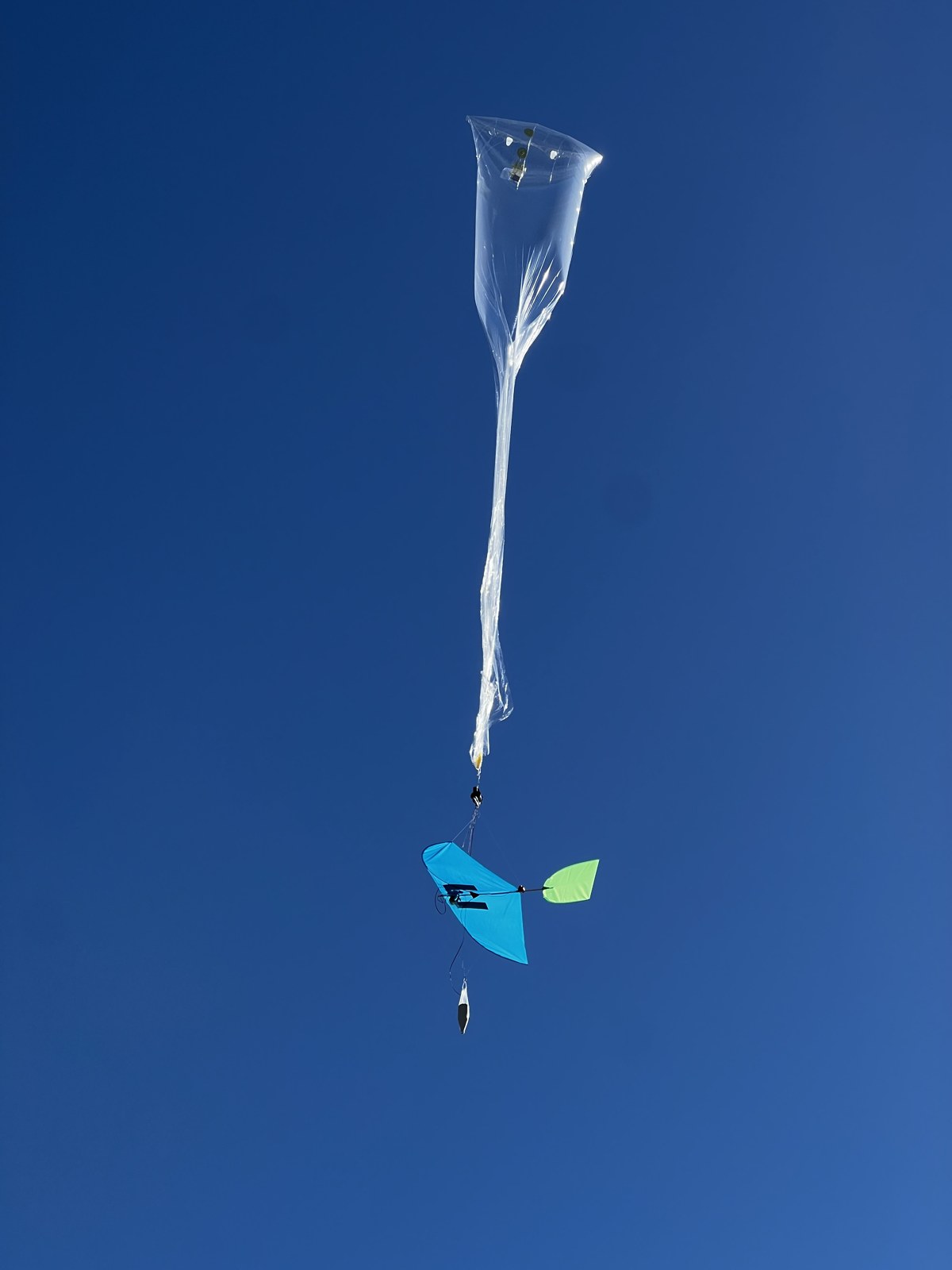}
  \caption{A Skysail system shortly after launch: a generation 2 wing with its tail and ballast hangs below a zero-pressure balloon. Later in flight the tether is deployed to \SI{450}{m}.}
  \label{fig:overview}
\end{figure}

\section{INTRODUCTION}

Vertical profiles of temperature, humidity and wind are the backbone of the global observing system. Today they come overwhelmingly from radiosondes: roughly \num{1300} upper-air stations release a balloon-borne sonde once or twice a day, on the order of \num{2000} launches per day worldwide~\lcite{durre2006igra,ecmwf2016radiosonde}. Each sonde ascends once, bursts after about two hours, and is discarded, so every profile costs a balloon, a sonde and a launch crew, and data collection is limited over oceans and the developing world.

Long-duration balloons replace the single ascent with a controllable platform capable of repeated profiling~\lcite{valbal_paper}. Altitude is controlled by venting lifting gas to descend and dropping ballast to ascend. By choosing its altitude the balloon also chooses its wind layer and can therefore navigate. Flights can last for weeks rather than hours, and flight durations of up to eleven months have been reported~\lcite{spectrum2026buoy}. This allows a single balloon to return tens of radiosonde-quality profiles~\lcite{spisak2026sensors}, which have been shown to improve global forecasts when assimilated~\lcite{windborne2023gdas}.

The value of such a balloon lies in the vertical distance it travels, not in the time it spends aloft. A balloon floating at constant altitude stays in the same pocket of air that it moves with and produces no new information. A balloon that repeatedly traverses the troposphere produces the profiles that meteorological assimilation systems are built around. Vertical motion also matters for the measurement itself: a balloon at float sits in nearly still air, so its sensors are poorly aspirated and suffer radiative and self-heating biases, whereas a balloon moving through the air continuously ventilates its sensors~\lcite{spisak2026sensors}.

However, every profile drops ballast that cannot be replaced. On a typical long-duration flight reversing the direction of a profile costs \SIrange{30}{40}{g} of lift, and the ballast supply of roughly \SI{1}{kg} is the single largest component of the launch mass and the quantity that ultimately ends the flight. The number of profiles a balloon can deliver is therefore bounded by the ballast it carries, and the cost per profile is bounded below by the cost of the hardware amortised over that number. Any mechanism that produces vertical motion without spending ballast or gas changes those economics: profiles become renewable, and the flight is limited by envelope life rather than by ballast.

This paper describes Skysail, an altitude-control system that extracts the work needed for vertical motion from the atmosphere itself (Fig.~\ref{fig:overview}): a small wing on a thin tether several hundred metres below the balloon, pitched in the wind shear between the two altitudes. We present the physics and sizing of the system, three generations of hardware, a deployment mechanism for the \SI{450}{m} tether, the control modes we flew, and the results of several dozen flight tests, including multi-day flights in which the balloon completed repeated profiles through the troposphere with essentially no ballast expenditure.

\section{RELATED WORK}

\subsection{Altitude control of free balloons}
A free balloon at equilibrium has no relative airflow to push against, so altitude control has traditionally meant changing free lift by venting gas or dropping ballast. 

Google Loon demonstrated pumping ambient air into a ballonet inside the helium envelope to descend and released it to ascend, giving fully reversible altitude control at the cost of a compressor working against a pressure differential that grows with altitude~\lcite{bellemare2020loon}. Loon balloons carried payloads of the order of \SI{100}{kg} with tens of square metres of solar panel. The compressor, its power system and the superpressure envelope that must contain the pumped air are heavy, and the approach does not scale down well to weather balloons that have to be light and cheap enough to be launched in large numbers. 

Propellers have also been proposed and flown for balloon trajectory control and for augmenting lift, including gossamer designs for the thin stratospheric air~\lcite{gossamer2024prop}. Vertical thrust is attractive because it is independent of the wind field, but the power required makes this infeasible for small, mass-deployable weather balloons, especially at high altitudes where the air density is lower.

\subsection{Extracting work from two fluids}
Skysail belongs to a family of machines that gain energy from the relative motion of two media. A sailboat is the canonical example: neither the air nor the water alone can drive it upwind, but a keel in one and a sail in the other can. The same principle allows a wind-driven cart to travel dead downwind faster than the wind~\lcite{drela2009ddwfttw}, lets albatrosses fly for days without flapping by cycling through the wind gradient over the sea surface~\lcite{sachs2005albatross}, and underlies crosswind kite power and kite propulsion of ships~\lcite{loyd1980crosswind,erhard2013kites}. In all these cases work is extracted from the velocity difference between the two media. For a balloon the two media are the air at two altitudes. The airborne-wind-energy literature treats the control of tethered wings in detail~\lcite{cherubini2015awe}, closed-loop dynamic soaring has been demonstrated on small aircraft~\lcite{bird2014soaring}, and the dynamics of a body on a long tether in wind are well studied for aerostats~\lcite{lambert2003aerostat}. Our system differs from all three above in that the tether is attached to a free-floating balloon rather to the ground, so the wing sees only wind shear, not absolute wind.

\subsection{Tethered surfaces below balloons}
Hanging an aerodynamic surface below a balloon to exploit wind shear has been studied before for \emph{horizontal} guidance. The StratoSail system hung a wing \SI{15}{km} below a stratospheric balloon, standing on edge so that its lift dragged the balloon sideways across the wind~\lcite{nock2007winged}, and Mazzoleni et al.\ analysed and flew a passive tethered sail for the same purpose, deriving its pitch stability and lateral force in closed form~\lcite{yoder2019sails,yoder2021tethered}. Lateral guidance is the natural goal for a superpressure balloon, which passively holds one altitude and cannot easily leave it. A zero-pressure balloon due to its flat lift profile (Eq.~\ref{eq:zplift}) can fly at any altitude, and by choosing its altitude against a forecast of the wind field it can achieve horizontal navigation. However, a low power and renewable mechanism for changing the altitude of a zero-pressure balloon remains an open problem, and is what this paper attempts to address. 

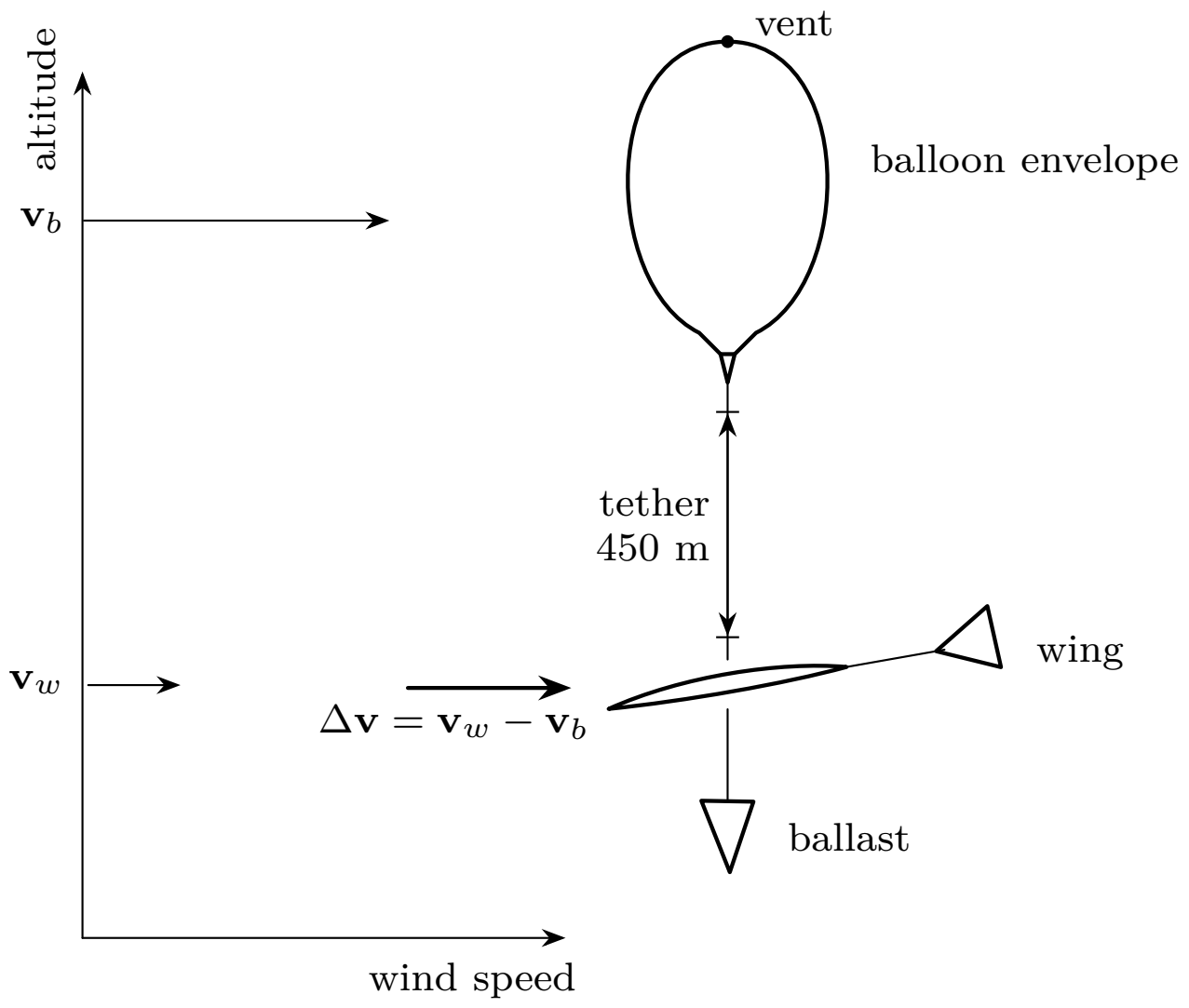
\begin{figure}[t]
  \centering
  \resizebox{\columnwidth}{!}{\begin{tikzpicture}[x=1cm,y=1cm,font=\footnotesize,>=Stealth,line cap=round,line join=round]
  \draw[->] (-2.04,-1.41) -- (-2.04,4.69);
  \draw[->] (-2.04,-1.41) -- (1.36,-1.41);
  \draw[->,thin] (-2.03,3.64) -- (0.12,3.64);
  \node[anchor=east,inner sep=2pt] at (-2.09,3.64) {$\mathbf{v}_b$};
  \draw[->,thin] (-2,0.37) -- (-1.35,0.37);
  \node[anchor=east,inner sep=2pt] at (-2.09,0.39) {$\mathbf{v}_w$};
  \draw[thick] (2.5,4.9) .. controls (3.4,4.9) and (3.4,3.2) .. (2.7,2.85) -- (2.55,2.7) -- (2.45,2.7) -- (2.3,2.85) .. controls (1.6,3.2) and (1.6,4.9) .. (2.5,4.9);
  \draw[thick] (2.45,2.7) -- (2.5,2.5) -- (2.55,2.7);
  \fill (2.5,4.9) circle (0.045); \node[anchor=south west,inner sep=1pt] at (2.65,4.9) {vent};
  \draw (2.5,2.5) -- (2.5,0.55);
  \draw[|<->|,thin] (2.5,2.3) -- (2.5,0.7) node[midway,left,align=right]{tether\\ 450 m};
  \begin{scope}[shift={(2.5,0.35)},rotate=10]
    \draw[thick,fill=white] (-0.85,0) .. controls (-0.3,0.14) and (0.4,0.12) .. (0.85,0) .. controls (0.4,-0.04) and (-0.3,-0.04) .. cycle;
    \draw (0.85,0) -- (1.55,0);
    \draw[thick] (1.9,0.25) -- (1.49,0) -- (1.92,-0.19) -- cycle;
    \draw[thick] (0.04,-0.82) -- (-0.32,-0.75) -- (-0.21,-1.28) -- cycle;
  \end{scope}
  \node[anchor=west] at (4.55,0.6) {wing};
  \draw (2.5,0.2) -- (2.5,-0.45);
  \node[anchor=west] at (2.8,-0.7) {ballast};
  \node[anchor=west] at (3.38,4.05) {balloon envelope};
  \node[anchor=west] at (-0.15,-1.71) {wind speed};
  \node[anchor=west, rotate=90] at (-2.33,3.86) {altitude};
  \draw[->,thick] (0.25,0.35) -- (1.4,0.35) node[pos=0.28, inner sep=3pt, below=0pt]{$\Delta\mathbf{v}=\mathbf{v}_w-\mathbf{v}_b$};
\end{tikzpicture}}
  \caption{The balloon--tether--wing system in a wind velocity gradient. The balloon drifts with the wind at its own altitude, so the wing \SI{450}{m} below sees the wind vector difference $\Delta\mathbf{v}$ as a relative airflow; its tail aligns it with $\Delta\mathbf{v}$ and its pitch sets the sign and magnitude of the vertical force transmitted up the tether.}
  \label{fig:physics}
\end{figure}

\section{METHOD}
\label{sec:method}

Skysail augments the two consumable actuators of a standard zero-pressure balloon with an aerodynamic one. The balloon envelope carries the gas envelope, and a LoRa radio-controlled vent actuator. Below it, on a \SI{450}{m} tether, hangs a wing that can control its pitch in relation to the gravity vector. A ballast container (with a ballast release actuator) hung beneath the wing provides the gravity vector reference and ensures the tether stays close to vertical (Fig.~\ref{fig:overview}). The wing module also contains an Iridium modem for global satellite connectivity, and a short-range LoRa link to command the remote vent. Wind speed and direction change with altitude, so the wing and envelope experience different wind vectors. As the envelope exhibits 1-2 orders of magnitude more sideways drag compared to the wing, it moves with the surrounding air, while the wing, carried at the balloon's horizontal velocity by the tether, experiences a relative airflow equal to the wind vector difference $\Delta\mathbf{v}$ between the two altitudes. The tail weathervanes the wing into that flow. Pitching the wing nose-up generates upward lift and the whole system rises, and nose-down generates downward force and it sinks. 

\subsection{Balloon Physics}
\label{sec:physics}

\subsubsection{Lift of a zero-pressure balloon}
A zero-pressure envelope is deliberately underfilled, so the lifting gas is equalized to the ambient pressure $p$ and, to a good approximation, ambient temperature $T$. For a mass $m_g$ of gas of molar mass $M_g$ in air of molar mass $M_a$ the gas volume is $V = m_g R T/(M_g p)$, with $R$ the gas constant, and the buoyant force is $\rho_a V g$ with the air density $\rho_a = M_a p/(R T)$, so the net lift
\begin{equation}
  L = \rho_a V g - m_g g = m_g g\left(\frac{M_a}{M_g} - 1\right)
  \label{eq:zplift}
\end{equation}
depends only on the gas mass and system mass, not on altitude. Below its ceiling (the altitude at which it becomes fully inflated) a zero-pressure balloon therefore has a fixed free lift $F = L - mg$, with $m$ the system mass, which sets the vertical motion. This independence of free lift from altitude allows a system with modest ability to trim the free lift to drive the balloon across a large range of altitudes.

\subsubsection{Free lift and vertical speed}
A free balloon moving vertically at speed $v_z$ experiences drag $D = \tfrac12 \rho_a v_z^2 C_D A$, where $A$ is the drag area of the envelope and $C_D$ its drag coefficient, so at terminal vertical speed
\begin{equation}
  v_z = \sqrt{\frac{2 F}{\rho_a C_D A}} .
  \label{eq:wvert}
\end{equation}
For the envelope used here ($A \approx \SI{0.5}{m^2}$, $C_D \approx 0.9$) a free lift of \SI{\pm 15}{g} gives $v_z \approx \SI{\pm 0.8}{m/s}$ at \SI{2}{km} and \SI{\pm 1.3}{m/s} at \SI{10}{km}, where the air is 2.4 times thinner, and a profile from \SIrange{2}{10}{km} takes a few hours. Reversing a profile with consumables means changing $F$ by $2|F|$, i.e. venting or dropping about \SI{30}{g} of lift per turnaround, consistent with the \SI{35}{g} per reversal spent by the ordinary flight of Fig.~\ref{fig:baseline}. A balloon carrying \SI{1}{kg} of ballast can therefore afford of the order of thirty ascents. Skysail replaces the change in $F$ with an aerodynamic force of the same order applied through the tether (Sec.~\ref{sec:flights}).

\begin{figure}[t]
  \centering
  \includegraphics[width=\columnwidth]{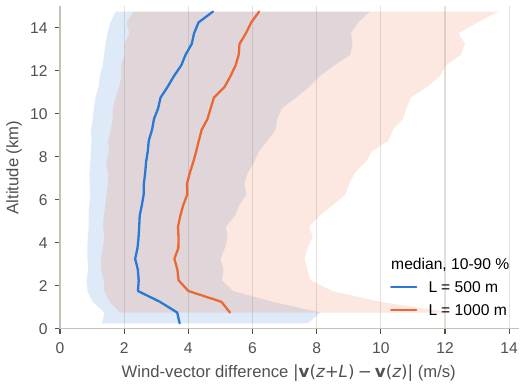}
  \caption{Wind shear available to a tethered wing, from GPS-derived winds on \num{8006} long-duration balloon profiles recorded in 2024--2025. Median and 10--90\% band of the wind vector difference across \SI{500}{m} and \SI{1000}{m} baselines versus altitude.}
  \label{fig:shear}
\end{figure}

\subsubsection{Aerodynamic force available from shear}
The wing lift is
\begin{equation}
  L_w = \tfrac12 \rho_a |\Delta\mathbf{v}|^2 C_L S_A .
  \label{eq:lift}
\end{equation}
Here $C_L$ is the lift coefficient and $S_A$ the wing area. With $C_L = 1$ and $S_A = \SI{0.6}{m^2}$ this is \SI{37}{gf} per \si{(m/s)^2} of shear at sea level and \SI{13}{gf} per \si{(m/s)^2} at \SI{10}{km}. The value $C_L = 1$ is a momentum bound at an angle of attack of \SI{30}{\degree}: a surface can at most remove all the momentum of the air passing through its projected area. The direct measurement of Sec.~\ref{sec:flights} (Fig.~\ref{fig:loaddelta}) is consistent with this bound. The force therefore scales with $\rho_a |\Delta\mathbf{v}|^2$, and a shear of \SIrange{1}{2}{m/s} across the tether is enough to profile in the troposphere.

\subsubsection{Tether length and the statistics of shear}
The force available depends on the wind difference across the tether. If the horizontal wind vector were a two-dimensional random walk in altitude, the expected $|\Delta\mathbf{v}|$ over a baseline $\ell$ would grow as $\sqrt{\ell}$~\lcite{chandrasekhar1943}, doubling the tether would buy only \SI{41}{\percent} more shear, and the difference would almost never vanish, because the probability density of a two-dimensional displacement is zero at the origin. In the limit of a prevailing wind in one direction at all altitudes the problem becomes one-dimensional and by Rolle's theorem any maximum of wind speed with altitude implies two levels with equal speed, and a wing spanning them sees no shear at all leading to an altitude band in which it would get stuck. The measured climatology (Fig.~\ref{fig:shear}) sits between the two limits: between \SI{2}{} and \SI{10}{km} the median $|\Delta\mathbf{v}|$ over a \SI{500}{m} baseline, close to the tether length, is \SI{2.6}{m/s}, and it falls below \SI{1}{m/s} at \SI{11}{\percent} of levels against \SI{35}{\percent} for the speed difference alone. We chose \SI{450}{m} (\SI{500}{yards}) as the standard tether for a practical reason: braided spectra fishing line is sold on \SI{500}{yard} spools, and a longer line costs altitude range, deployment reliability and risk of snapping in high wind shear.

\subsection{Hardware Design}
\label{sec:hardware}

\begin{figure}[t]
  \centering
  \IfFileExists{figures/photo_gen1_kite.jpg}{\includegraphics[height=0.42\columnwidth]{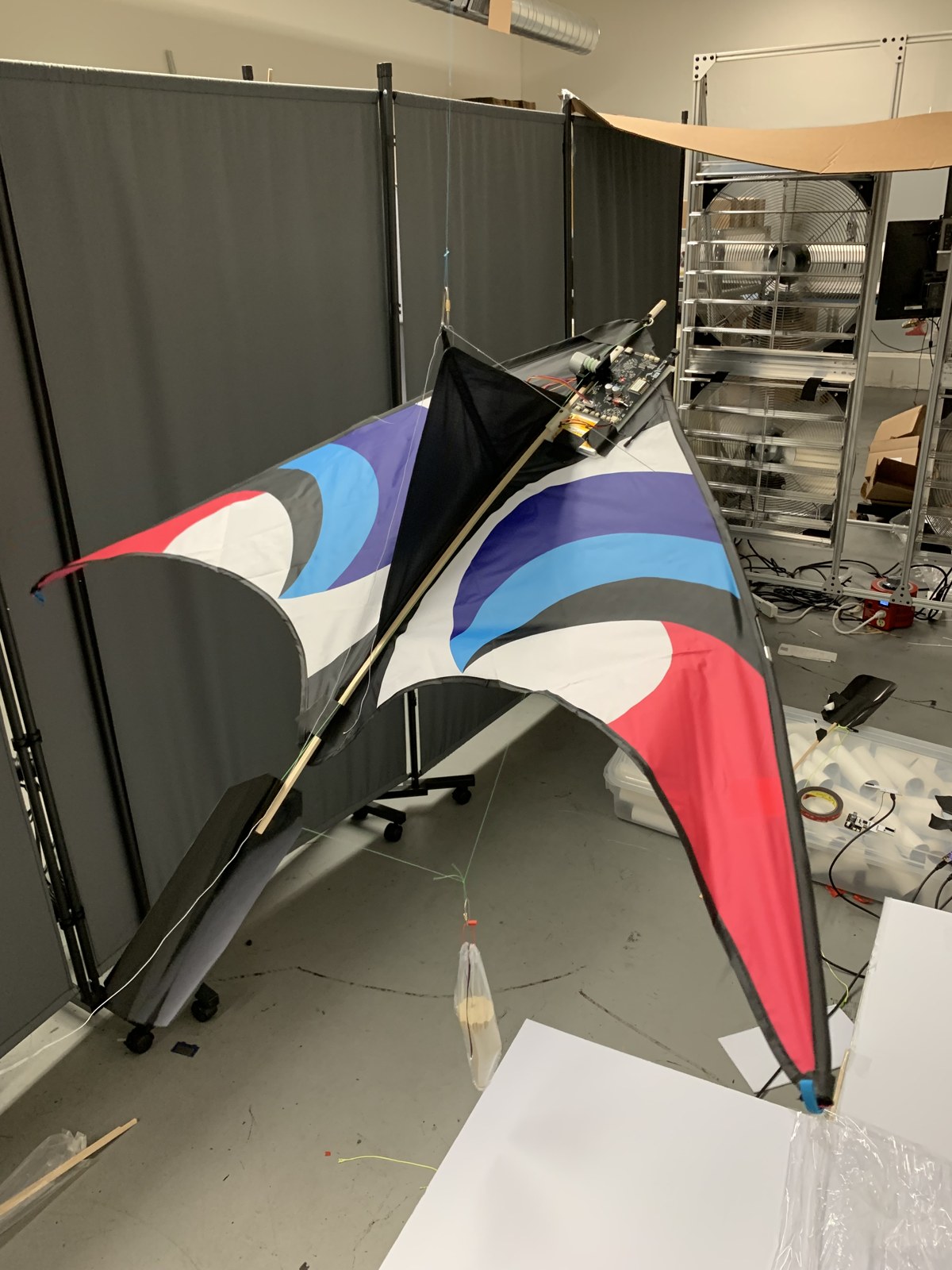}}{}\hfill
  \IfFileExists{figures/photo_gen2_sewn.jpg}{\includegraphics[height=0.42\columnwidth]{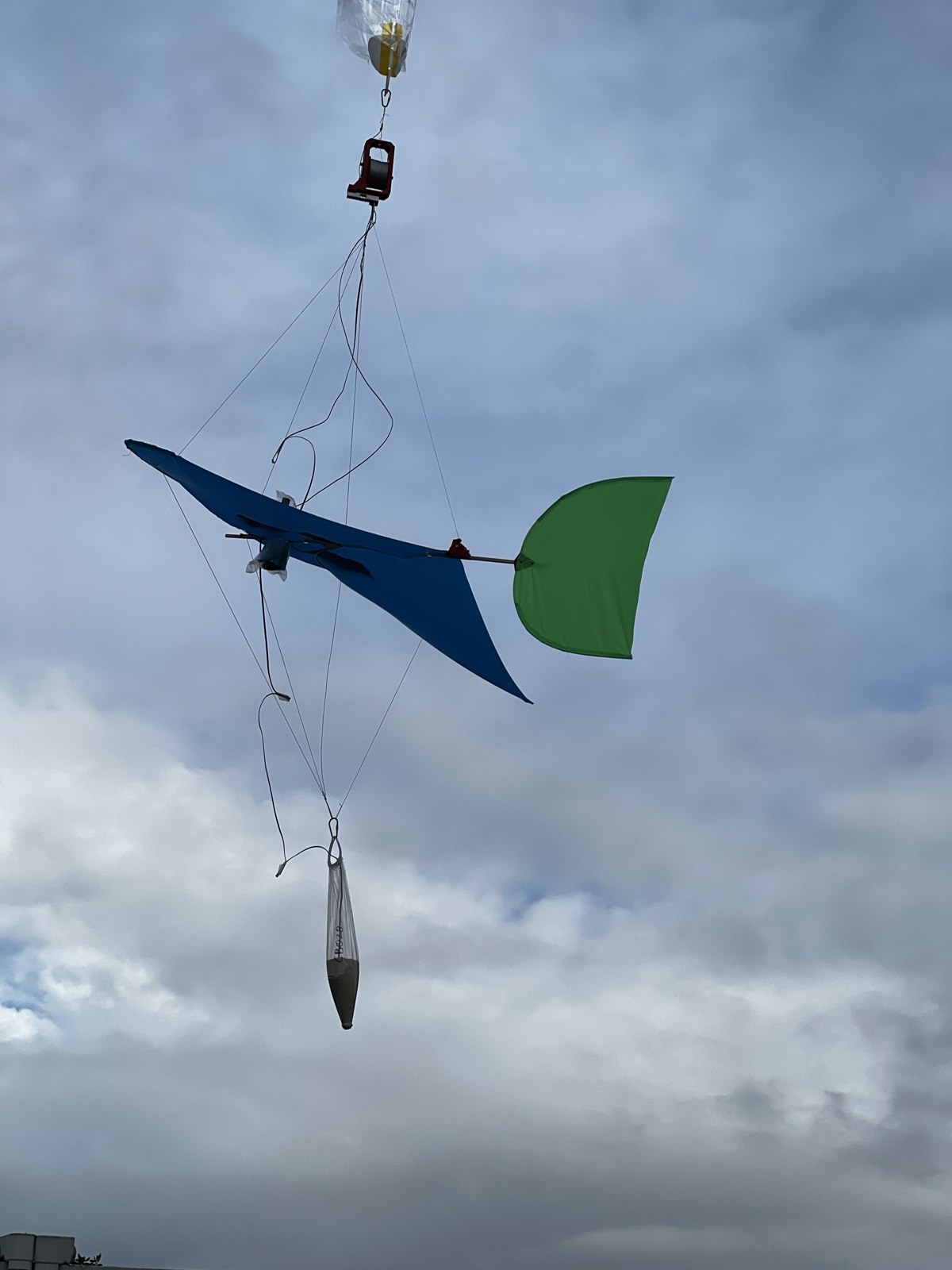}}{}\hfill
  \IfFileExists{figures/photo_gen3_pe.jpg}{\includegraphics[height=0.42\columnwidth]{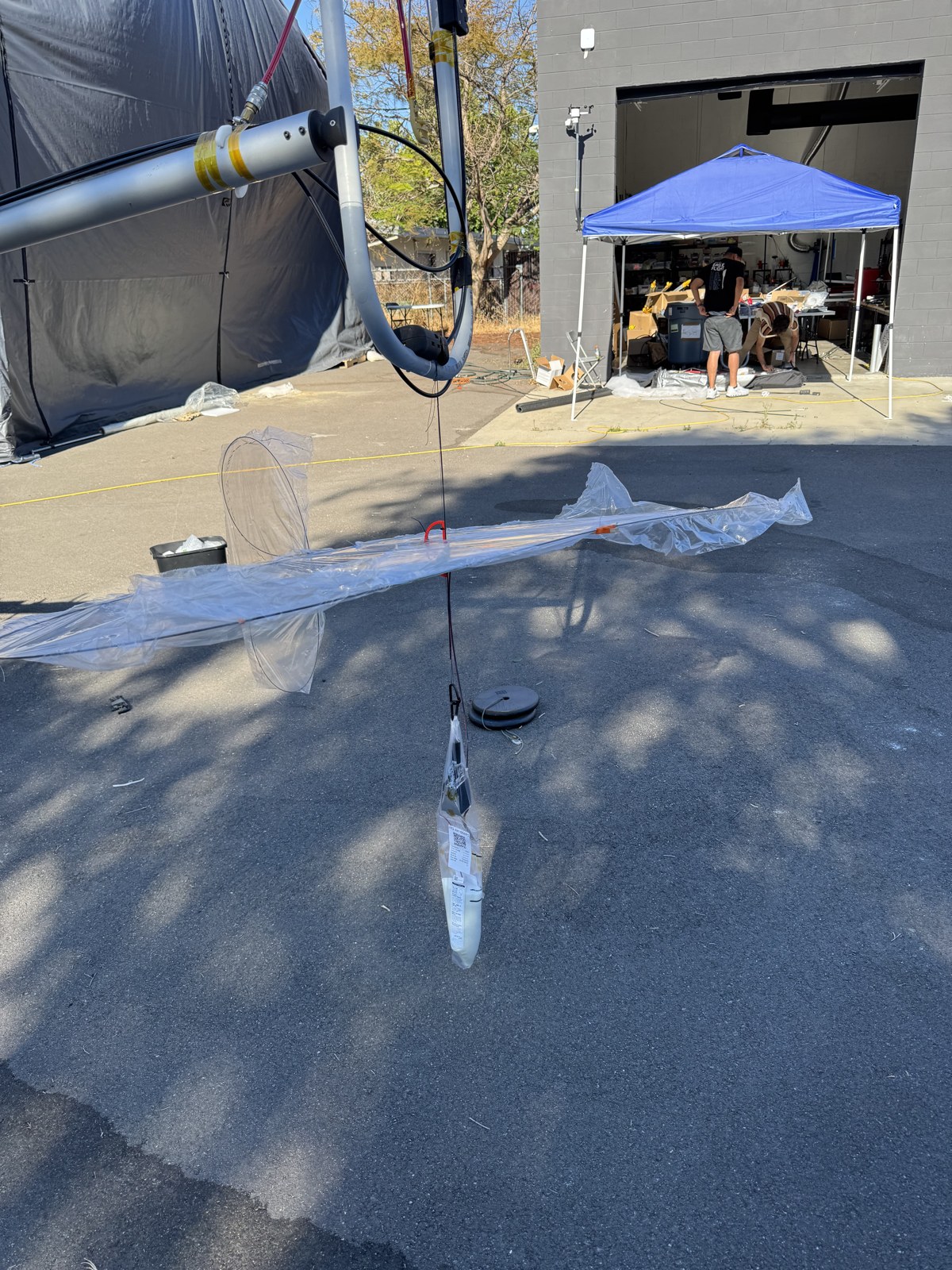}}{}
  \caption{Three generations of sail hardware. Left: an off-the-shelf kite with pitch set by a weight moved along a line beneath it. Centre: a ripstop nylon wing with carbon rib spars and string-driven pitch in flight. Right: heat-sealed polyethylene wing with a worm-gear pitch drive and a transparent tail, shown before launch.}
  \label{fig:generations}
\end{figure}

Every generation of the sail module shares the same architecture (Fig.~\ref{fig:generations}): a lifting surface of roughly \SI{0.6}{m^2}, a pitch actuator, an inertial measurement unit, a \SI{915}{MHz} LoRa radio and optional load cell mounted in various places for experiments. 

\paragraph{Generation 1 (kite)} An off-the-shelf kite was equipped with a custom electronics board. Pitch was set by moving a weight along a line running fore and aft beneath the kite, a small funicular driven by a motor and spool on the electronics board, shifting the centre of gravity relative to the tether attachment point and giving a new angle of attack. While this was enough to show that the system produces a measurable and reversible vertical force it showed a stock kite did not generate enough lift variation in slow winds, and the weight-shift trim system didn't hold the kite stable enough.

\begin{figure}[t]
  \centering
  \resizebox{0.95\columnwidth}{!}{\begin{tikzpicture}[x=1cm,y=1cm,font=\footnotesize,>=Latex]
  \draw[thick] (5.02,3.2) -- (5.01,2.4);
  \draw[thick, blue] (4.19,1.08) -- (5.03,-0.4);
  \draw[thick, purple] (5.86,1.39) -- (5.03,-0.4);
  \node[above, node font=\scriptsize] (node2) at (5,3.22) {tether to balloon};
  \coordinate (T) at (5.02,2.4);
  \draw[thick, blue] (T) -- (4.19,1.08);
  \node[anchor=east, align=right, draw=none, text=blue, node font=\scriptsize] at (4.51,0.2) {fixed lines\\ (leading edge)};
  \node[anchor=west, align=left, text=purple!80!black, node font=\scriptsize] (node1) at (5.25,2.5) {control line\\ (trailing edge)};
  \draw[thick, purple] (5.02,2.4) -- (5.86,1.39);
  \node[anchor=west, node font=\scriptsize] at (5.32,-0.66) {ballast};
  \node[anchor=west, node font=\scriptsize] at (6.42,0.99) {tail};
  \draw[<->,thick] (5.94,1.8) arc (40:-20:0.9);
  \begin{scope}[shift={(143pt,35pt)}, rotate=10]
      \draw[thick, fill=yellow] (-0.85,0) .. controls (-0.3,0.14) and (0.4,0.12) .. (0.85,0) .. controls (0.4,-0.04) and (-0.3,-0.04) .. cycle;
      \draw (0.85,0) -- (1.55,0);
      \draw[thick, fill=yellow] (1.9,0.25) -- (1.49,0) -- (1.92,-0.19) -- cycle;
      \draw[thick] (-0.11,-1.65) -- (-0.47,-1.58) -- (-0.36,-2.11) -- cycle;
    \end{scope}
  \draw[fill=red] (5.85,1.38) circle (0.07cm);
\end{tikzpicture}}
  \caption{Pitch actuation of the generation 2 sail. Fixed lines run from the tether point to the leading edge; a control line runs to the trailing edge and is wound on a motorised spool (red circle) on the trailing edge of the wing; winding it "walks" the trailing edge along the cables. The ballast hangs from the module.}
  \label{fig:actuation}
\end{figure}
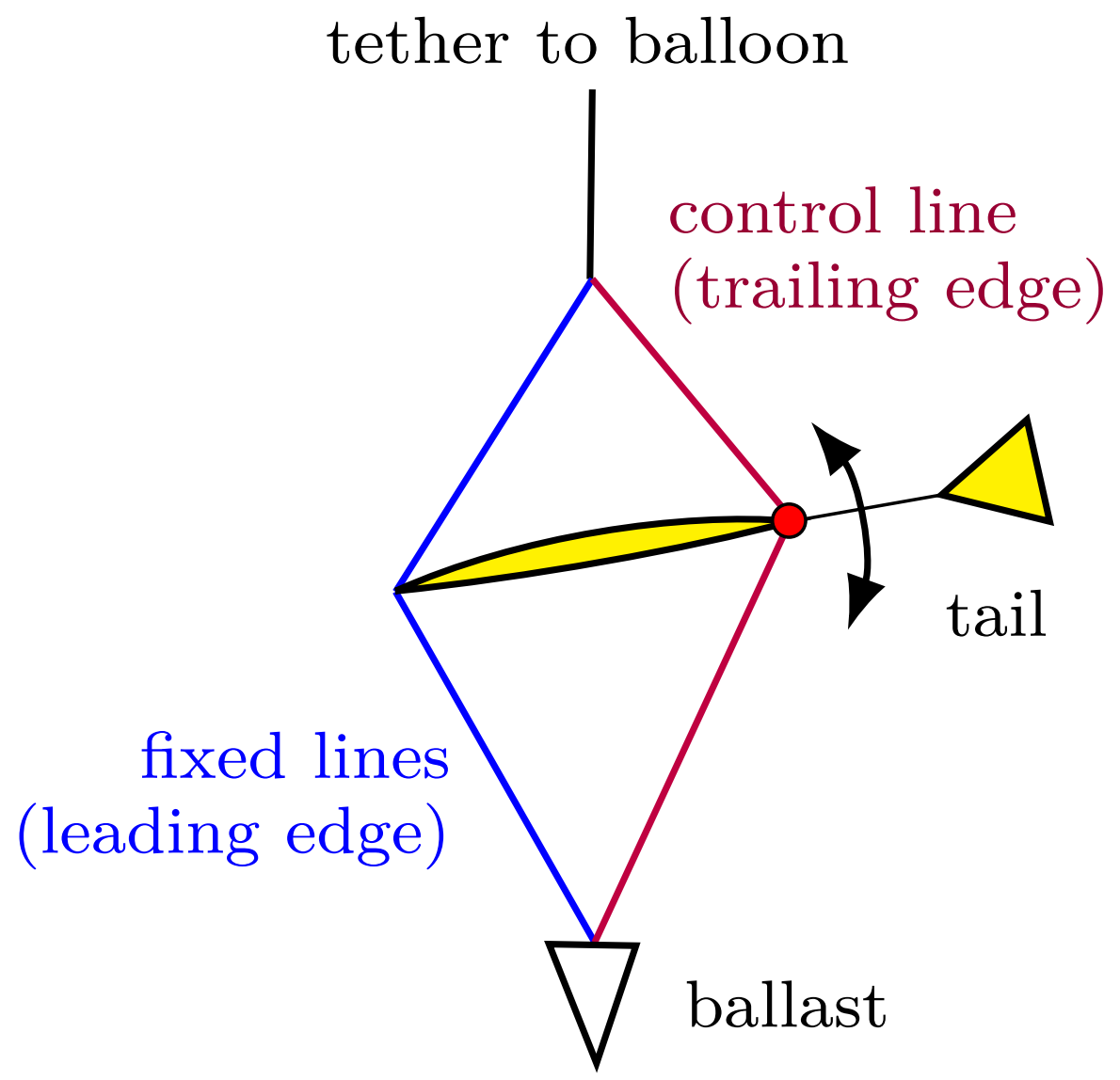

\paragraph{Generation 2 (custom wing)} A custom ripstop nylon wing allowed us to get more lift in the low wind-shear regime and a more reliable pitch mechanism. Carbon-fibre rib spars are sleeved into the fabric, which is cut on a fabric plotter and sewn with curved hems so that the skin is tensioned by the spars. Pitch is set by a spool on the electronics board that rides along a control line to the trailing edge against fixed lines to the leading edge, so the wing rotates about its leading-edge attachment (Fig.~\ref{fig:actuation}). Having multiple leading-edge lines provides roll stability. Placing the point of rotation so far forward also allowed the system to return to zero angle of attack during large gusts preventing damage. This generation produced most of the results of Sec.~\ref{sec:flights}.

\paragraph{Generation 3 (design for manufacturing)} The third generation wing is a rectangle heat-sealed from polyethylene film stretched over arched carbon rods, avoiding the need to sew. Another issue we wanted to eliminate is the difficulty with assembling the many control lines of generation 2. The generation 3 pitch mechanism has no untensioned lines. The diameter of a geared semicircle is held vertical using the mass of the ballast. The wing is mounted to the semicircle using a rotational joint and has a worm gear that engages with the geared semicircle to rigidly change its pitch. Additionally this allows the platform's standard, unmodified zero-pressure balloon payload to hang under the sail acting as its electronics. 

\paragraph{Remote vent} Because the balloon's vent is at the top of the envelope and the flight computer now sits several hundred metres below it, the vent is driven by its own small radio board at the top of the envelope, powered by a coin cell and a solar patch connected over a \SI{915}{MHz} LoRa link. It vents on command, reports its barometric altitude, which doubles as the measurement of tether length, and carries a deadman timer that vents the balloon down if the link is lost.

\subsection{Deployment Mechanism}
\label{sec:deploy}

A \SI{450}{m} tether cannot be laid out on the ground, so the sail is launched with its tether wound on a spool, and is lowered to its working length in flight. Deployment turned out to be the hardest engineering problem in the project and the dominant cause of lost flights. The difficulty is energetic: lowering the roughly \SI{1}{kg} sail-and-ballast module through \SI{450}{m} releases about \SI{4.5}{kJ}, which must be dissipated by a mechanism that weighs a few grams, and the line must pay out without snarling from a package that has just been handled, shaken and accelerated at launch.

\begin{figure}[t]
  \centering
  \includegraphics[width=0.8\columnwidth]{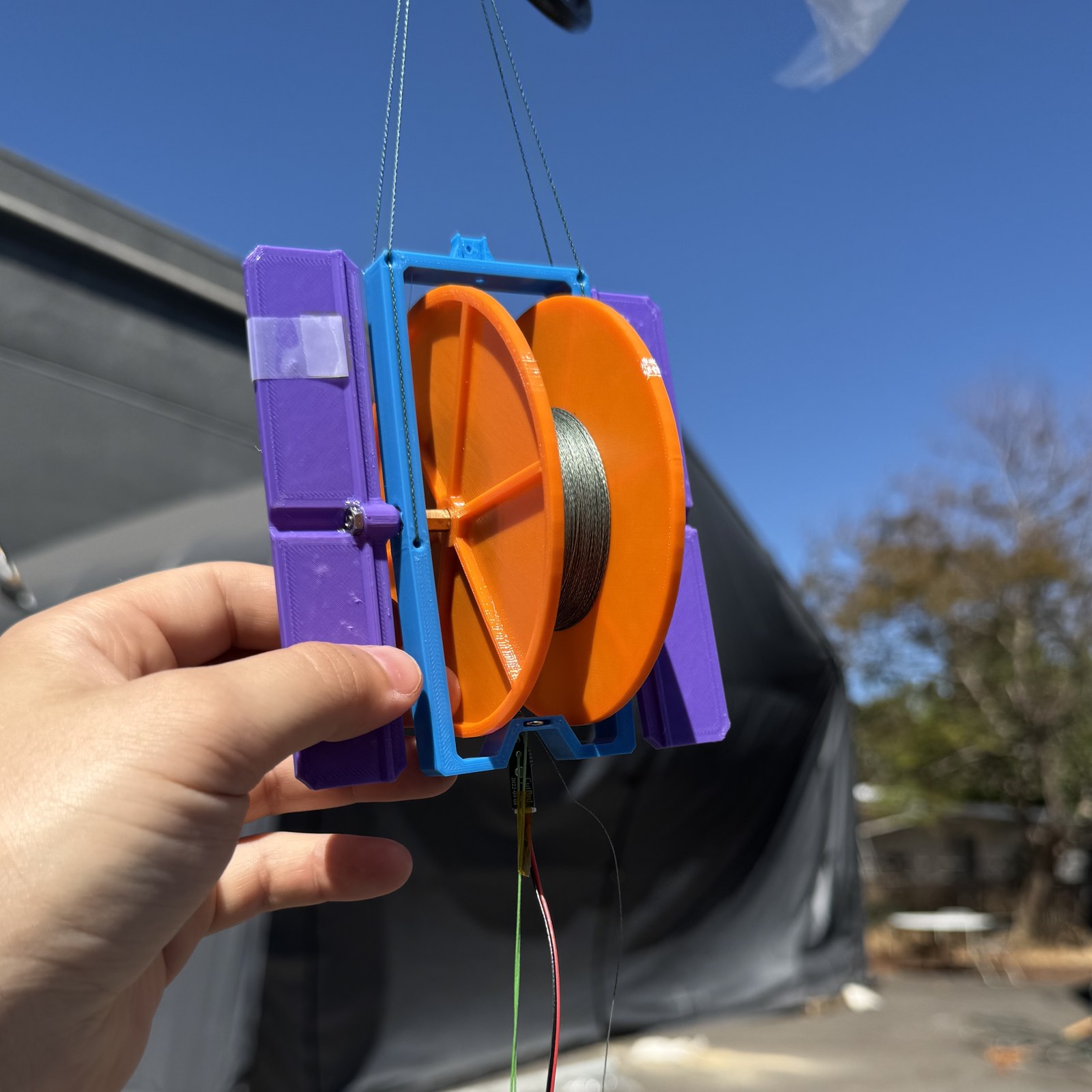}
  \caption{The aerodynamic governor. The line is wound around the hub of a 3D-printed spool; the two paddles are the air brake, and the shrinking winding radius slows the payout as the line runs out (Eq.~\ref{eq:governor}).}
  \label{fig:governor}
\end{figure}

\begin{figure}[t]
  \centering
  \includegraphics[width=\columnwidth]{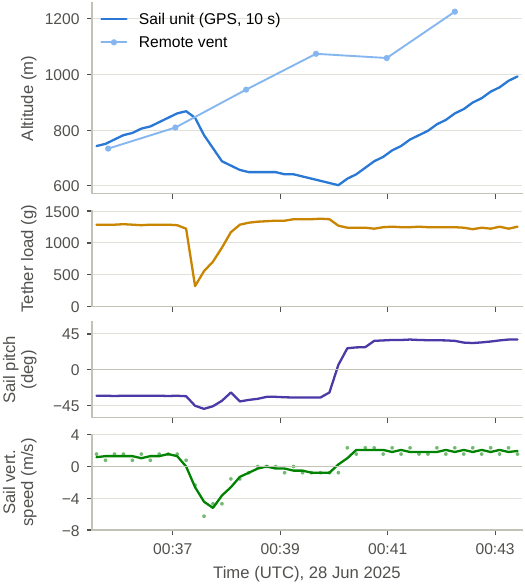}
  \caption{An in-flight tether deployment (F-3327, 28 June 2025). From top: GPS altitude of the sail module and barometric altitude of the remote vent, which separate as the line pays out (the remote vent altitude has a lower temporal resolution and lags because it is relayed over an extra radio hop); tether load at the sail; sail pitch; sail vertical speed. The gradual re-loading of the line after the drop is the governor at work.}
  \label{fig:deploy}
\end{figure}

After prototyping several schemes we converged on an aerodynamic governor (Fig.~\ref{fig:governor}). The line is wound around the hub of a 3D-printed spool with two paddles on its faces that act as an air brake whose torque grows with the square of the rotation rate $\omega$, $\tau_b = c\,\omega^2$, where $c$ is set by the paddle geometry. As the line pays out the spool spins, and the falling module of mass $m$ pulls the line with tension $T \approx m g$, applying a torque $T r$ at the current winding radius $r$, so at steady state $T r = c\,\omega^2$ and the payout speed $u$ is
\begin{equation}
  u = \omega r = \sqrt{\frac{m g\, r^3}{c}} \;\propto\; r^{3/2}.
  \label{eq:governor}
\end{equation}

The governor therefore slows the payout automatically as the spool empties: the winding radius shrinks from its full value $r_0$ to the hub radius $r_1$, and the speed falls by the factor $(r_1/r_0)^{3/2}$. Our spool is \SI{80}{mm} in diameter when full and \SI{6}{mm} at the hub, so Eq.~\ref{eq:governor} predicts that the governor alone would slow the payout by a factor of $(80/6)^{3/2} \approx 50$ by the time the last turn leaves the hub. In practice the balloon is climbing, so the payout cannot fall below the balloon's own ascent rate and the wing's terminal velocity, but the governor remains important as it applies tension to the line preventing tangling. Fig.~\ref{fig:deploy} shows the telemetry of a deployment with an inline load cell measuring the tension in the tether. The sail's vertical speed peaks at about \SI{-6}{m/s} against the balloon's \SI{+1.5}{m/s} climb, and the gradual re-loading shows the governor limiting the speed when near the end of the line. The altitude difference between the remote vent and the sail confirms the deployment.

The tether is braided ultra-high-molecular-weight polyethylene fishing line of \SIrange{20}{30}{lb} nominal strength, chosen to stay below the \SI{50}{lb} break-strength limit that keeps the system exempt from FAA Part 101. Knots reduce the strength of this line, and tensile tests gave a conservative \SIrange{15}{20}{lb} for the end-to-end system.

\subsection{Control Methods}
\label{sec:control}

The system is controlled by three nested loops: an inner loop that holds the wing at a commanded pitch, a middle loop that chooses which pitch to command when the system wants to go up or down, and an outer loop that decides when it wants to go up or down.

\paragraph{Inner loop} The sail module estimates its pitch from the accelerometer and gyroscope of its inertial measurement unit with a complementary filter~\lcite{mahony2008complementary}. The pitch actuator is driven by an on--off controller with hysteresis~\lcite{astrom2021feedback}: the motor runs at a fixed duty cycle toward the setpoint whenever the error exceeds a threshold and stops once it falls inside a smaller one. The expanded deadband when stationary prevents motor chatter as gusts on the wing move the estimate by a few degrees.

\paragraph{Middle loop} We flew three schemes for choosing the pitch setpoint.
\begin{enumerate}
  \item \emph{Fixed pitch} One nose-up angle for going up and one nose-down angle for going down, typically \SI{\pm 30}{\degree}. This ignores the change in apparent wind from the system's own vertical motion, which in practice mattered little.
  
  \item \emph{Sweep for extreme tension} With the load cell inline with the tether, the pitch is stepped through its range in \SI{5}{\degree} increments with a \SI{10}{s} dwell at each, then set to the angle that gave the largest tension (to go down) or the smallest (to go up), and the sweep is repeated every \SI{20}{min} as the shear changes with altitude.
  
  \item \emph{Compliant control} A load cell in series between the wing and the pitch spool carries the control-line tension, which, once the wing's weight at each pitch is calibrated out, measures the aerodynamic torque about the pivot and hence indirectly the angle of attack relative to the airflow. Stepping the pitch by \SI{5}{\degree} whenever this tension leaves a set band holds the wing at a roughly fixed angle to the airflow rather than to gravity.
\end{enumerate}

Somewhat surprisingly, the more complex schemes gave little advantage, and we predominantly used the fixed pitch scheme, which is also tolerant of unreliable load-cell measurements. 

\begin{figure}[t]
  \centering
  \includegraphics[width=\columnwidth]{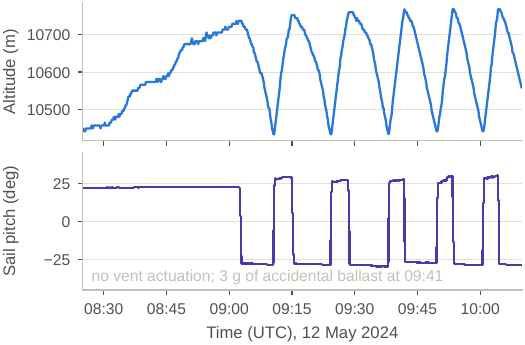}
  \caption{Holding a precise altitude with small sail-only profiles: F-1296 at \SI{10.5}{km} on 12 May 2024. A \SI{\pm 30}{\degree} pitch square wave kept the balloon within a narrow altitude band for multiple cycles.}
  \label{fig:w1296saw}
\end{figure}

\paragraph{Outer loop} 
The system commands a change of direction on its own in two situations. The first is leaving an altitude band set from the ground. The commanded reversal drives the balloon back into the band, and because the request is held until the opposite edge is reached, the balloon bounces between the two levels (Fig.~\ref{fig:w1296saw}). The second is when the standard vent-and-ballast controller predicts a violation of its safety bounds. It then asks the sail first, and falls back on the vent or ballast only if the shear across the tether has died out and the sail cannot arrest the motion. This fallback accounts for the ballast spent on the multi-day flights of Sec.~\ref{sec:flights}.

\begin{figure}[t]
  \centering
  \includegraphics[width=\columnwidth]{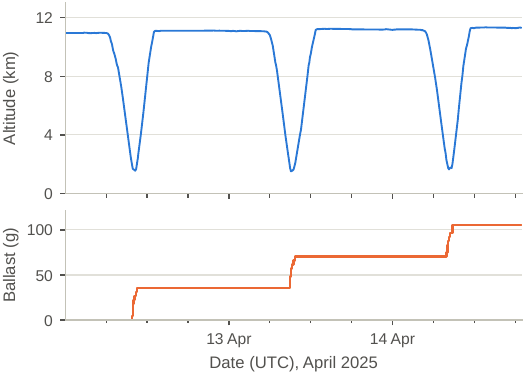}
  \caption{Baseline: three days of an ordinary flight of the platform without a sail (F-2684, 12--14 April 2025). Top: altitude. Bottom: ballast dropped. Each of the three reversals at the bottom of a profile is bought with an average of \SI{35}{g} of ballast.}
  \label{fig:baseline}
\end{figure}

\begin{figure}[t]
  \centering
  \includegraphics[width=\columnwidth]{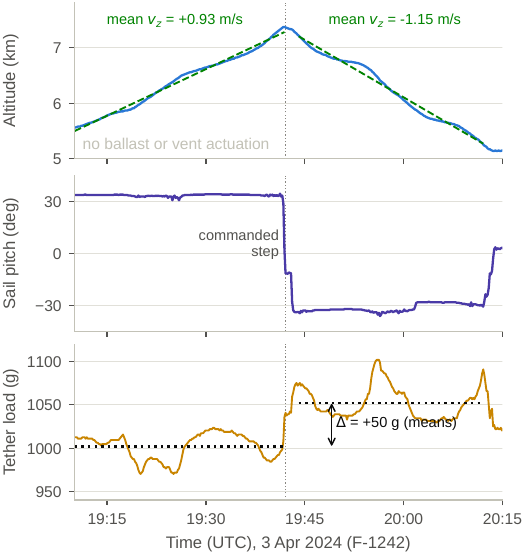}
  \caption{Direct measurement of the sail force on F-1242 (3 April 2024, \SI{7.3}{km}). Top: altitude, with the mean vertical speed over the pre-command and post-command holds. Middle: measured sail pitch with the commanded reversal. Bottom: tether load with its mean over the same two periods.}
  \label{fig:loaddelta}
\end{figure}

\section{FLIGHT TESTS}
\label{sec:flights}

Between August 2023 and August 2025 we launched \num{42} skysail balloons accumulating approximately \num{97} balloon-days of recorded flight and \SI{2900}{km} of vertical travel. Of these, \num{24} flights lasted longer than a day Fourteen of the \num{42} launches produced a working sail; the rest were lost to deployment (10), tether or attachment (5), balloon or envelope (5), control, electronics or communications (7) faults, or launch damage (1). The flights presented below are representative of the standard configuration unless noted. Fig.~\ref{fig:baseline} shows a normal flight of the platform without a sail, for comparison of ballast usage and profile shape.

\subsection{Direct measurement of the aerodynamic force} 
The load cell in line with the tether measures the force the sail applies to the balloon. Fig.~\ref{fig:loaddelta} shows an example of such a measurement, on flight F-1242: a commanded pitch step from \SI{+30}{\degree} to \SI{-30}{\degree}, held for half an hour on either side. Averaged over the two holds, the tether load rose by \SI{50}{g} and the vertical speed went from \SI{+0.9}{} to \SI{-1.2}{m/s}. These forces are of the size the balloon's own drag requires. With $A = \SI{0.5}{m^2}$, $C_D = 0.9$ and the air density at \SI{7.3}{km} ($\rho_a = \SI{0.57}{kg/m^3}$), Eq.~\ref{eq:wvert} gives an envelope drag of \SI{11}{gf} at \SI{+0.9}{m/s} and \SI{17}{gf} at \SI{-1.2}{m/s}, so reversing the motion needs about \SI{28}{gf}. The \SI{50}{g} measured on the tether is of that order, the excess being consistent with the drag of the sail module itself. 

\subsection{Altitude hold using small profiles} The same authority can hold the balloon at a precise altitude. Fig.~\ref{fig:w1296saw} shows F-1296 using a \SI{\pm 30}{\degree} pitch square wave to stay within a \SI{330}{m} altitude band without using consumables, while still moving through the air fast enough to ventilate its sensors.

\begin{figure}[t]
  \centering
  \includegraphics[width=\columnwidth]{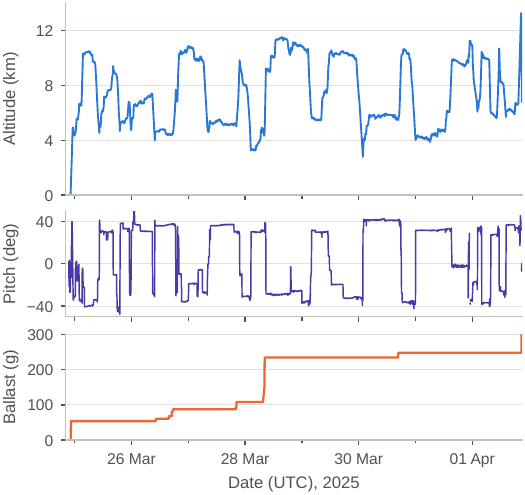}
  \caption{Eight-day flight F-2581 with a larger (\SI{1}{m^2}) wing. Top: altitude. Middle: sail pitch. Bottom: ballast dropped. The large drop on 28 March was a controller issue (Sec.~\ref{sec:flights}). Near the end strong wind shear caused the tether to break, and the step at the end is the flight computer dumping ballast during the fall.}
  \label{fig:w2581}
\end{figure}

\begin{figure}[t]
  \centering
  \includegraphics[width=\columnwidth]{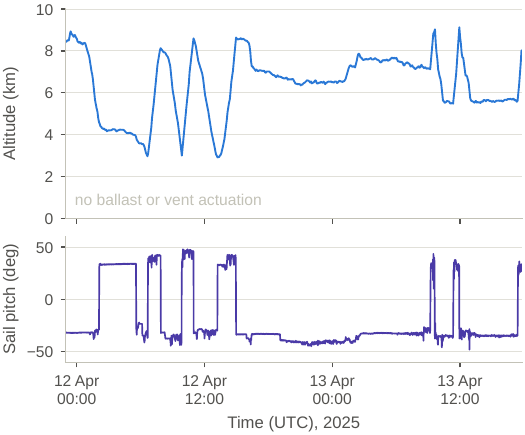}
  \caption{Long profiles flown with the sail alone: F-2673, 11--13 April 2025. Top: altitude. Bottom: measured sail pitch.}
  \label{fig:w2673}
\end{figure}

\subsection{Ballast-free profiling} 

Fig.~\ref{fig:w2673} shows 43 hours of flight in which eleven reversals, \SI{53}{km} of vertical travel, were flown with no ballast or vent actuation at all. In the middle of the record there was not enough wind shear anywhere in the altitude range to drive the system down, and it sat between \SI{6}{km} and \SI{8}{km}. Waiting a day allowed it to resume profiling without expending ballast.

\subsection{Long duration flight} 
Fig.~\ref{fig:w2581} shows the full history of an eight-day flight. The balloon completed 24 profiles of more than \SI{1}{km} and \SI{159}{km} of vertical travel. Of the \SI{248}{g} of ballast dropped before the terminal descent, \SI{127}{g} went in a single controller episode on 28 March in which an automated ground system commanded the balloon to climb faster than the sail could drive it. Additionally \SI{54}{g} of ballast went into post launch trim. Excluding those the flight used \SI{67}{g}, about \SI{3}{g} per profile turnaround against \SI{35}{g} for the ordinary flight of Fig.~\ref{fig:baseline}. The residual ballast was spent almost entirely at the bottoms of profiles, offsetting lift gas that leaked from the balloon. Between those episodes the counter is flat for a day at a time.

\begin{figure}[t]
  \centering
  \includegraphics[width=\columnwidth]{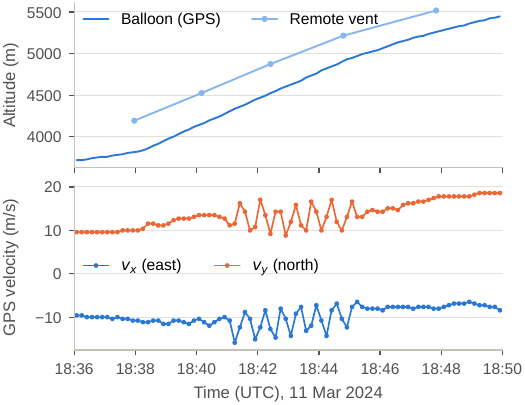}
  \caption{Tether pendulum motion on F-1178. Top: balloon altitude and the altitude of the remote vent, whose separation grows as the line pays out. Bottom: east and north GPS velocity of the wing, oscillating with the \SI{40}{s} period of a \SI{400}{m} pendulum ($T_p = 2\pi\sqrt{\ell/g}$).}
  \label{fig:pendulum}
\end{figure}

\subsection{Pendulum behaviour} The tethered system is a \SI{450}{m} pendulum. Its \SI{40}{s} period $T_p$ occasionally appears in the wing module's GPS velocity (Fig.~\ref{fig:pendulum}). The oscillation decays within tens of minutes and does not affect control, however it pollutes the balloon's wind measurement, which is derived from GPS velocity, and it must be filtered before the winds can be used in meteorological applications. It also provides an additional measurement of deployed tether length, $\ell = g(T_p/2\pi)^2$.

\section{CONCLUSIONS AND FUTURE WORK}

We have shown that a small wing on a \SI{450}{m} tether can control the altitude of a zero-pressure balloon using the wind shear across the tether, and that in the troposphere this control is strong enough to carry a balloon through repeated profiles for days at a time with the consumable actuators nearly idle. A pitch command reverses the balloon's vertical motion within minutes, and the sail alone can hold altitude to stay in a desirable wind field or cycle altitude to collect meteorological data. The resource the system exploits is available most of the time in most of the troposphere, and where it is not, the balloon simply waits or falls back on ballast. 

Several directions follow directly from these flights. The controller is currently blind to the shear it depends on. A forecast wind field would let the balloon choose altitudes where the shear across the tether is strong and avoid the levels where it vanishes, and the same forecast could be used to plan whole profiles rather than react to bounds. Where a single tether length finds no shear, two or more wings stacked on a longer tether would span more of the wind profile and reduce the chance that every wing sits in a null at once. The flights reported here were on envelopes whose gas leakage set their own ceiling on duration. A low-leak envelope would extend the number of profiles a flight can deliver.

Finally, the continuous vertical motion that the sail makes possible also ventilates the sensors, reducing the impact of radiative and self-heating on biases in the temperature and humidity measurements, however this remains to be quantified for meteorological applications.

\section*{ACKNOWLEDGMENT}

The research, experiments, and analysis in this paper were conducted without the use of AI tools. Claude Fable 5.1 (Anthropic) was used to draft the paper from detailed author-written outlines, and to generate the plotting code for Figs.~\ref{fig:shear}, \ref{fig:deploy}, \ref{fig:loaddelta}, \ref{fig:w1296saw}, \ref{fig:w2581}, \ref{fig:baseline}, \ref{fig:w2673} and~\ref{fig:pendulum} and the diagrams of Figs.~\ref{fig:physics} and~\ref{fig:actuation} from author-provided data.  All AI-drafted text and figures were reviewed, edited, and approved by the authors, who take full responsibility for the content.

\ifanon\else
The Skysail system, its deployment and its control laws are the subject of a patent by the authors~\lcite{sushko2025skysail}. We thank WindBorne Systems for supporting this work.
\fi

\bibliographystyle{IEEEtran}
{\raggedright\bibliography{refs}}  %

\end{document}